# Quenching the Non-Collinear Spin Order in High-$T_c$ Layered Ferromagnet Fe$_5$GeTe$_2$

Rabindra Basnet and Ramesh C. Budhani[*]

*Department of Physics, Morgan State University, Baltimore, MD, 21251, USA*

## Abstract

The realization of long-range spin order in two-dimensions (2D) has catapulted the search for layered materials with magnetic ordering above room temperature. These efforts aim to understand and enhance the spin–spin interactions in 2D. An emergent class of such magnets is the layered Fe$_N$GeTe$_2$ ($N$ = 3, 4, and 5). Here, we investigate the magnetic states over a wide field–temperature phase space in the high-$T_c$ ferromagnet Fe$_5$GeTe$_2$ using magnetization, ferromagnetic resonance (FMR), and magneto-transport measurements. Our findings reveal a magnetic phase transition from a collinear to a complex non-collinear magnetic order near the temperature $T^* \approx 160$ K, below which magnetic susceptibility is reduced, FMR linewidth broadened, and anomalous Hall resistivity suppressed. Such non-collinearity results from the competition between magnetocrystalline anisotropy and Dzyaloshinskii–Moriya interaction arising from the unusual Fe1 ordering in two possible split sites. Our study focuses on the strategy to quench the non-collinear spin order. Substituting 40% Ni in Fe$_5$GeTe$_2$ is found to be one such quenching strategy. This provides deeper insights into the magnetism of a high-$T_c$ layered-ferromagnet, offering opportunities to develop 2D magnet-based devices.

[*]ramesh.budhani@morgan.edu

## I. INTRODUCTION

Two-dimensional (2D) magnetism has attracted much attention due to fundamental scientific interest in low-dimensional magnetic orders and the potential use of 2D magnets in spintronics. Recent studies have led to the discovery of a wide range of van der Waals (vdW) materials where the magnetic order survives down to the 2D limit [1–5]. Such groundbreaking works greatly expand the realm of 2D magnetism and provide material platforms for investigating low-dimensional magnetism. However, most vdW magnets order well below room temperature and have low electrical conductivity. These factors have hindered their applicability in magneto-electronic devices under ambient conditions. Thus, new room-temperature itinerant vdW magnets that generate strong spin-polarized currents are highly desired for next-generation technology.

In the 2D limit, the Curie temperature $(T_c)_{2D}$ can be estimated as $(T_c)_{2D} = (T_c)_{3D}/\ln[3\pi(T_c)_{3D}/4K]$ [6]. Here, $K$ is the anisotropy constant, and $(T_c)_{3D}$ is the $T_c$ in three dimensions given by $(T_c)_{3D} = 4\pi J/N$, where $J$ is the measure of magnetic exchange coupling, and $N$ is the number of components. Usually, the anisotropy, which is mainly driven by spin-orbit coupling (SOC) and crystal field effects, is relatively weaker than the exchange interaction in 2D vdW magnets, resulting in much smaller $T_c$ in the 2D limit as compared to the $T_c$ of 3D ferromagnets [6,7]. Such a fundamental limitation imposes a challenge to designing high-$T_c$ vdW ferromagnets. One model magnetic material of the vdW family that can overcome this challenge is $Fe_NGeTe_2$ (FGT). So far, only three stable phases of FGT corresponding to $N = 3$ ($Fe_3GeTe_2$), $N = 4$ ($Fe_4GeTe_2$), and $N = 5$ ($Fe_5GeTe_2$) have been found [8–10]. Unlike the other vdW ferromagnets [2,3,5], some FGT compounds exhibit $T_c \geq 300$ K down to a unit cell thickness [4,8,11]. The high ordering temperature in these materials is achieved by creating a network of magnetic elements, which forms stronger 3D-like spin-pair interactions within the 2D vdW structure [8]. In addition to room-temperature ferromagnetism, the FGT compounds exhibit metallic conductivity, providing a convenient route to access exotic functional properties arising from the interplay between magnetism and electronic transport. Various scientifically interesting and technologically useful phenomena, such as the large topological Hall effect (THE) [12,13], topological nodal line semimetal [14,15], large anomalous Nernst effect [16], magnetic skyrmion [17,18], etc., have been observed in this system. Thus, the FGT materials are potential candidates for room-temperature spintronic devices.

Besides practical applications, the magnetic properties of FGT are also intriguing from a fundamental point of view. For example, $Fe_3GeTe_2$ is known to host magnetic skyrmions and related spin textures [17,18]. Similarly, the sister compound $Fe_4GeTe_2$ features spin reorientation at low temperatures, offering a platform to explore the fundamentals of magnetic domains, skyrmion evolution, and phase transitions in 2D magnets. Compared to $Fe_3GeTe_2$ and $Fe_4GeTe_2$, the incorporation of an additional layer of Fe in $Fe_5GeTe_2$ enhances the $T_C$. However, it also introduces greater complexity to magnetic order. For instance, a wide range of $T_c$ between $T_c \approx$ 270 to 332 K has been reported in this material [10,11,19–21] together with the sensitivity of magnetic properties to thermal cycling history [19,22]. The magnetism in $Fe_5GeTe_2$ is strongly coupled with the crystal lattice. A recent work has demonstrated spin reorientation due to the alteration of vdW stacking order [19], while another study suggests a helimagnetic order arising from inversion symmetry breaking of the crystal [23]. At low temperatures, $Fe_5GeTe_2$ also exhibits unusual behavior, hosting multiple magnetic phases [23,24], distinct magnetic domains, and spin textures, including skyrmions [25]. Such magnetic anomalies at low temperatures are unique in a high-$T_c$ vdW ferromagnet and thus demand an in-depth investigation for a clear understanding of their origin.

Here, we have investigated the magnetic properties of $Fe_5GeTe_2$ using magnetization, ferromagnetic resonance (FMR), and electrical transport measurements. These probes reveal anomalies at low temperature, which are attributed to the transition from a collinear FM state to a complex non-collinear spin structure below $T^* \approx 160$ K. Our findings suggest the competition between magnetocrystalline anisotropy and the antisymmetric Dzyaloshinskii-Moriya interaction (DMI) as the physical origin of such non-collinear magnetic order. This mechanism is further clarified by quenching the non-collinear magnetic phase via 40% Ni-substitution in $Fe_5GeTe_2$. Our work provides a deeper understanding of magnetic phase evolution in vdW-type materials, which could be useful for tuning the spin structure in 2D magnets.

## II. EXPERIMENT

The single crystals of $Fe_5GeTe_2$ used in this work were synthesized by a two-step synthesis method. First, a polycrystalline precursor was prepared by heating the stoichiometric mixture of Fe (99.98%, Thermo Scientific), Ge (99.999%, Thermo Scientific), and Te (99.99%, Thermo Scientific) at 950°C for 24 hours. The source chemicals were sealed in an evacuated (pressure ~$10^{-5}$ Torr) quartz tube using an oxygen-acetylene torch. The precursor was finely grounded and then used as the source for chemical vapor transport (CVT) using iodine as the transport agent. As shown in Fig. S1a in the supplementary material [26], the millimeter-sized single crystals with flat and shiny surfaces were obtained after 10 days of CVT growth with a temperature gradient from 800 to 700°C. These crystals were characterized by X-ray diffraction (XRD) at room temperature using a Rigaku miniflex diffractometer with Cu-$K\alpha$1 radiation. A typical diffraction pattern of the crystals placed such that the scattering vector is parallel to the c-axis has only the (00$L$) reflections as shown in Fig. S1b [26]. Furthermore, a similar growth method has been used to synthesize the single crystals of Ni-substituted $Fe_5GeTe_2$, i.e., $(Fe_{1-x}Ni_x)_5GeTe_2$, where $x$ = 0.15, 0.25, and 0.40, which were characterized using the room-temperature XRD presented in Fig. S1b [26]. Ni atoms have been found to preferentially substitute at the Fe1 sites in F5GT [27]. The magnetization and electrical transport measurements were performed in a physical property measurement system (PPMS EverCool, Quantum Design). For the measurements of ferromagnetic resonance (FMR), a custom-built cryo-FMR setup [28], which operates in the frequency modulation mode has been used. Here, the sample was placed on the U-shaped signal line of a grounded co-planar waveguide where the frequency-modulated GHz input from an RF source is fed, and the rectified return signal is measured using a lock-in detection technique. The mutually perpendicular static and RF magnetic fields are in the sample plane (i.e., *ab*-plane).

**III. RESULTS AND DISCUSSION**

**Low-temperature non-collinear spin texture in Fe$_5$GeTe$_2$.**

The magnetic properties of Fe$_5$GeTe$_2$ (F5GT) are highly sensitive to sample growth conditions [10] because of the presence of multiple Fe sites in the unit cell whose occupancy depends on temperature, its gradient across source and sink along the tube, cooling method (natural cooling or quenching), choice of the source precursors, and the transport agent used during growth [10,11,22,29]. Therefore, we have adopted a cautious two-step synthesis method. First, the polycrystal precursor is prepared, which is finely grounded and used as a source along with I$_2$ as a transport agent for single-crystal growth. This method yielded large, plate-like single crystals with hexagonal facets and metallic luster. An optical microscope image of a F5GT single crystal is shown in the supplementary material [26] (Fig. S1a) with a *c*-axis lattice parameter of $c \approx 28.84$ Å, consistent with previous reports [10,11,27]. FGT crystallizes in a rhombohedral structure (space group $R\bar{3}m$) [10,19] with each unit cell consisting of three identical layers separated by a vdW gap and made up of 2D slabs of Fe and Ge sandwiched between two Te layers as shown in the inset of Fig. S1c [26]. Such layer stacking is often termed as ABC stacking [11,19], where the center Ge atoms are surrounded by three different Fe sites, i.e., Fe1, Fe2, and Fe3. The Fe1 atoms situated in the outermost plane of the Fe$_5$Ge subunit possess two possible split-sites above or below the Ge atom. Such intricate Fe1 ordering is crucial in determining the symmetry of the crystal and its magnetic properties [23,30]. To emphasize this distinction, Fe1 atoms are represented as partially filled blue spheres in the crystal structure diagram (Inset, Fig. S1c [26]), in contrast to the solid blue spheres used for the Fe2 and Fe3 sites. Compared to Fe$_3$GeTe$_2$ and Fe$_4$GeTe$_2$, magnetism in F5GT is richer. Earlier studies of this compound have predicted a complex non-collinear magnetic order [30], including helimagnetism at low temperatures [23]. To investigate this phenomenon, the temperature dependence of susceptibility ($\chi$) has been measured with in-plane ($H//ab$; $\chi_{ab}$) and out-of-plane ($H//c$; $\chi_c$) magnetic fields under a zero-field-cooling (ZFC) protocol, shown as hollow and solid circles in Fig. 1a, respectively. Here, the $\chi_{ab}$ and $\chi_c$ are measured at 0.1 (purple color) and 1 tesla (blue color), which are below and above the critical fields for magnetization saturation, respectively (as will be discussed later). As shown in Fig. 1a, the paramagnetic (PM) to FM transition temperature ($T_c$) is ≈ 316 K, which agrees well with the reported values [10,11,19,23,24]. Below the $T_c$, $\chi_{ab}$ for a 0.1 tesla first increases before showing a downturn at temperatures below $T ≈ 160$ K. The temperature near which $\chi_{ab}$ at 0.1 tesla starts to decrease is hereafter assigned as $T^*$, which is precisely determined by the peak position in the derivative d$\chi$/d$T$, as depicted in the inset of Fig. S2 [26]. This susceptibility behavior featuring a broad hump near $T^* ≈ 160$ K is not clear in the 0.1 tesla out-of-plane field measurement ($\chi_c$). Such low-temperature magnetic anomalies associated with competing magnetic orders in F5GT may be sensitive to ZFC and field-cooled (FC) susceptibility measurements [29]. Therefore, we performed the in-plane susceptibility measurements under both ZFC and FC protocols at various values of applied fields *ranging from* 0.002 to 1 tesla as shown in Fig. S2 [26]. These measurements demonstrate that the broad hump is present in both protocols for fields between $\mu_0H$ = 0.002 tesla and 0.5 tesla, indicating that the ZFC and FC conditions do not significantly impact this anomaly. For clarity, we present only the ZFC susceptibility data at $\mu_0H$ = 0.1 and 1 tesla fields in Fig. 1a. The characteristic temperature $T^*$ for each field is determined from the derivative d$\chi$/d$T$, as marked by the purple arrows in the insets of Fig. S2 [26], corresponding to the dashed lines in the susceptibility curves. The $T^*$ is found to be strongly dependent on the applied field. Increasing the field above 0.5 tesla suppresses this feature. The magnetic anomaly in $\chi_{ab}$ near $T^*$ and its disappearance with increasing magnetic field above 1 tesla has been reported earlier [23,24]. The vanishing $T^*$ at higher fields can be attributed to the stabilization of a collinear spin arrangement along the field direction. The field dependence of $T^*$ is summarized in Fig. 1b. Notably, the anomaly persists up to $\mu_0H$ = 0.5 tesla at $T^* ≈ 134$ K and shifts monotonically to higher temperature with decreasing magnetic field, reaching $T^* ≈ 190$ K at 0.002 tesla. The evolution of $T^*$ at even lower magnetic fields remains unknown due to the intrinsic limitation of magnetization measurements, which require a finite applied field to generate a measurable signal. The persistence of this magnetic anomaly at or near zero field, versus its stabilization solely under applied magnetic fields, remains an open question. Therefore, alternative experimental techniques such as neutron scattering are needed to probe the low- or zero-field magnetic states in this system. Multiple mechanisms, such as magnetic ordering of Fe1 sublattice [10], emergence of helimagnetism [23], non-unidirectional magnetic moments between adjacent F5GT layers [24], and FM to ferrimagnetic (FIM) transition [20], have been suggested to explain the origin of this anomaly at low field. Furthermore, a recent theoretical work has predicted the presence of a complex non-collinear magnetic order at low temperature in F5GT [30]. Indeed, this temperature $T^* \approx 160$ K is critical in Fe5GT. Spin reorientation accompanied by modified magnetic domains has also been observed near $T^*$ in magnetic force and scanning tunneling microscopy studies [22]. In addition to the anomalies near $T_c$ and $T^*$, the $\chi_{ab}$ measured at 0.1 tesla also exhibits an additional feature near $\approx 50$ K under both ZFC (denoted by a red triangle in Fig. 1a) and FC protocols (supplementary material [26]; Fig. S2f). This anomaly, consistently reported in earlier studies under similar temperature and field conditions, has been primarily attributed to a first-order magneto-structural transition [10,11,19,29], with the additional possibility of a separate magnetic phase transition [23].

To further understand the magnetic anomaly at $T^*$, the field dependence of magnetization [$M(B)$] from $T = 340$ to 10 K under in-plane ($H//ab$) and out-of-plane ($H//c$) magnetic fields is shown in Fig. 1c as solid and dashed lines, respectively. Below $T_c$, the isothermal magnetization exhibits field-induced saturation along both directions for the entire temperature range, as expected in ferromagnets. The critical field for magnetic saturation for in-plane magnetization $(H_s)_{ab}$ is smaller than the out-of-plane magnetization $(H_s)_c$ at all temperatures. This behavior indicates persistent in-plane magnetic anisotropy over the entire temperature range below $T_c$ and confirms the absence of any temperature-driven switching of the magnetic easy axis. The in-plane easy axis for F5GT is distinct compared to the out-of-plane easy axis for Fe$_3$GeTe$_2$ [31] and temperature-driven spin reorientation between the $ab$-plane and $c$-axis below $T_c$ for Fe$_4$GeTe$_2$ [8]. Additionally, the F5GT sample exhibits the unusual variation of saturation fields with temperature compared to Fe$_3$GeTe$_2$ [31] and Fe$_4$GeTe$_2$ [8]. As seen in Fig. 1d (upper panel), the $(H_s)_{ab}$ and $(H_s)_c$ display non-monotonic evolution below and above $T^* \approx 160$ K. The critical fields first reduce upon cooling from 300 to 220 K, and then become nearly constant down to $\approx 160$ K. A further drop in temperature enhances both, but the enhancement of $(H_s)_{ab}$ is larger than that of $(H_s)_c$. This results in the suppression of magnetic anisotropy, causing a nearly-isotropic magnetic order at $T < 160$ K.

Such non-monotonic evolution of $H_s$ and tuning of anisotropy occurring below $T < 160$ K suggests a magnetic phase transition near $T^* \approx 160$ K, as depicted by different colored regions in Fig. 1d, where the ferromagnetic state is marked as FM and NC and a non-collinear (NC) magnetic order, which will be discussed later.

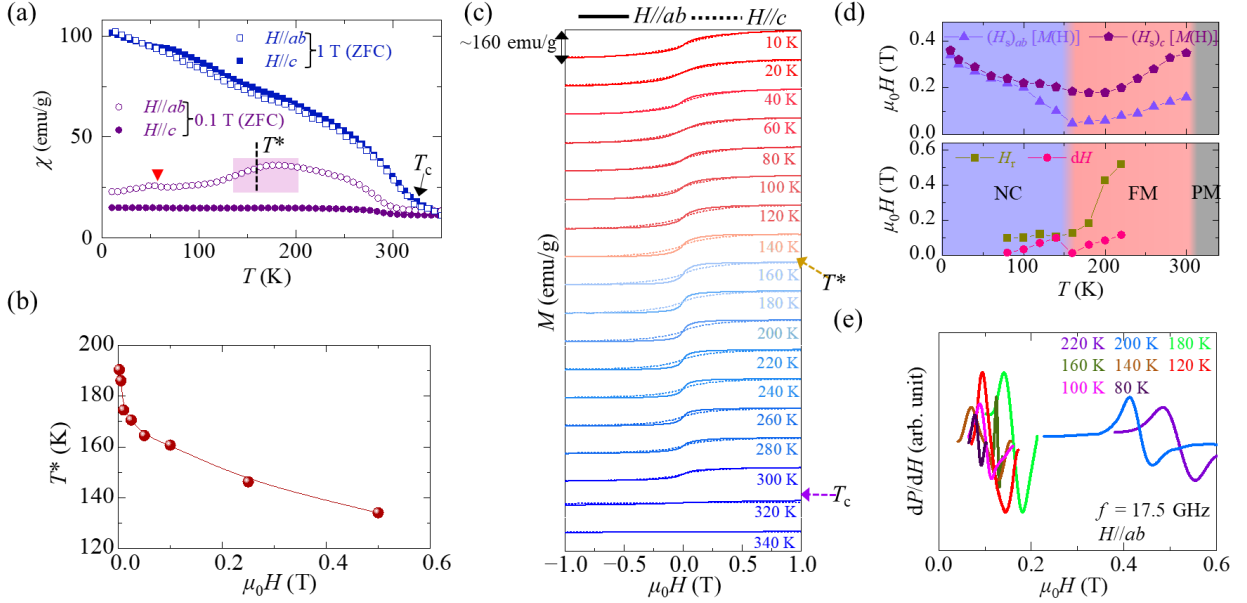

*Fig. 1. Magnetic properties of Fe$_5$GeTe$_2$. (a) Temperature dependence of susceptibility ($\chi$) for pristine Fe$_5$GeTe$_2$ samples for H//ab and H//c magnetic fields of 0.1 tesla (Purple color) and 1 tesla (Blue color) under a ZFC protocol. The black arrow denotes the PM to FM transition temperature $T_c \approx 316$ K. The highlighted region (Rectangle; magenta color) depicts the magnetic anomaly, which appears around $T^* \approx 160$ K (dashed line) for a 0.1 tesla field. (b) Field dependence of $T^*$ extracted from the in-plane susceptibility measurements presented in Fig. S2. (c) The M(H) for pristine Fe$_5$GeTe$_2$ samples for H//ab (solid line) and H//c (dashed line) from $T = 340$ K to 10 K. (d) Variation of in-plane [$(H_s)_{ab}$] and out-of-plane [$(H_s)_c$] critical fields for magnetization saturation with temperature (upper panel). The $(H_s)_{ab}$ and $(H_s)_c$ fields are extracted from the M(H) measurements presented in Fig. 1c. The different colored regions in Fig. 1d represent different magnetic phases, which will be explained later in the main text. They delineate the transition from the high-temperature PM to the FM state near $T_c \approx 316$ K, as well as the onset of NC spin order below $T^* \approx 160$ K. (e) FMR spectra of a single crystal of Fe$_5$GeTe$_2$ from $T = 220$ to 80 K. The FMR $H_r$ and dH for each temperature are presented in Fig. 1d (lower panel).*

The modification of the magnetic state in the vicinity of $T^*$ is further established through the FMR measurements performed at several temperatures close to $T^*$ with a 17.5 GHz excitation under an external DC magnetic field directed along the *ab*-plane of the sample. The broadband FMR spectroscopy of F5GT has been widely studied at room temperature [32–34]. On the other hand, the FMR behavior of this material at low temperatures remains relatively unexplored. An earlier study has shown that the FMR signal of the F5GT crystal is barely detectable and randomly scattered with varying input frequencies at cryogenic temperatures [20]. The FMR spectra of F5GT down to $T = 80$ K are presented in Fig. 1e and used to extract the resonance field ($H_r$) and the linewidth (d$H$) of the resonance by fitting them with a two-component Lorentzian function,

$$\frac{dP}{dH} = K_1 \frac{4dH(H-H_r)}{[4(H-H)^2+(dH)^2]^2} - K_2 \frac{(dH)^2-4(H-H_r)^2}{[4(H-H_r)^2+(dH)^2]^2}, \quad (1)$$

where $K_1$ and $K_2$ are the symmetric and anti-symmetric coefficients, and $H$ is the applied DC field. The $H_r$ and d$H$ obtained from the fitting at all temperatures are summarized in Fig. 1d (lower panel). The $\mu_0 H_r \approx 0.52$ tesla at $T = 220$ K agrees well with the reported values near this temperature [20]. Both $H_r$ and d$H$ reduce on cooling down to $\approx 160$ K, which is consistent with the suppression of $(H_s)_{ab}$ between 300 to 160 K, as discussed above. The $H_r$ only changes slightly on reducing the temperature below 160 K, whereas d$H$ suddenly increases, followed by a monotonic decrease with temperature. The broadening of the linewidth on cooling from 300 to 150 K was also observed in a previous FMR study of the bulk F5GT sample, which has been ascribed to FM to FIM transition [20]. The two-magnon scattering activated by defects can also broaden the FMR linewidth in ultrathin ferromagnets [35]. However, the two-magnon contribution is usually weak in bulk single crystals due to fewer structural defects. Indeed, the high $T_c$ ($\approx 316$ K) observed in our F5GT sample, which is comparable to $T_c \approx 317$ K reported for the stoichiometric composition [24,27], together with the fact that the $T_c$ in FGT compounds is markedly suppressed by structural defects [30,36], indicates negligible vacancies in these samples. Thus, the $H_r$ and d$H$ variations seen in Fig. 1d indicate the magnetic phase tuning near $T^* \approx 160$ K.

The itinerant magnetic nature of FGT compounds enables electrical readout of the spin states [37]. The magnetic phases strongly influence charge carrier scattering in materials, and thus magnetic transitions are reflected in the measurements of electrical resistivity. As seen in Fig. 2a, the longitudinal resistivity of the F5GT crystal exhibits a distinct change in slope in the vicinity of $T_c$ and $T^*$, the two critical temperatures derived from the susceptibility data of Fig. 1a. The residual resistivity ratio (RRR) calculated from the data of Fig. 2a is estimated to be approximately ~8. The higher RRR of our F5GT sample is in line with the near-stoichiometric composition and highlights its superior crystalline quality compared to the Fe-deficient sample studied in an earlier work, where RRR was only ~1.4 [10]. We also performed the measurements of magnetoresistance (MR) defined as MR = $(\rho(H)-\rho(0))/\rho(0)$, where $\rho(H)$ and $\rho(0)$ are the longitudinal resistivity of the sample in a field $H$ applied perpendicular to the plane of the crystal and in zero-field, respectively. The behavior of MR between 340 and 10 K is shown in Fig. 2b. The MR is negative over the entire temperature range, which is expected for a ferromagnet where the electron-magnon scattering is suppressed under applied field. Again, consistent with magnetization and FMR results, the MR displays distinct behavior in the two magnetic regimes separated by $T^*$ as marked in the figure by the yellow line. A normal negative MR is detected above $T^*$ (Fig. 2b). In contrast, for $T < T^*$, a sudden drop in MR occurs near the out-of-plane saturation field $(H_s)_c$ extracted from $M(H)$ measurements (Fig. 1c), which is depicted by the dashed lines in Fig. 2b. Therefore, the field-dependent MR in the $T < T^*$ regime can be divided into two regimes of field, suggesting distinct magnetic states of the crystal. In the low-field regime indicated by the dashed lines (Fig. 2b), the observed behavior is associated with the non-collinear (NC) magnetic order. As the magnetic field increases beyond these lines, MR drops significantly due to the progressive alignment of the non-collinear spins, which reduces spin-disorder scattering. Once the system reaches the FM state at higher fields, most spins are already aligned with the field, so further suppression of spin-disorder scattering is minimal, resulting in only a slight change in MR. Hence, the high-field MR regime is represented as the FM state. Figure 2e (upper panel) summarizes the behavior of MR as a function of temperature at the saturation field of 6 tesla. In the PM state, the $|MR|_{6T}$ slightly increases on cooling from 340 to 320 K. On entering the FM state, the $|MR|_{6T}$ shows a slight drop at 300 K, leading to a small kink near $T_c$ that implies enhanced magnetic scattering in the vicinity of the PM-FM transition. On cooling below $T \approx 300$ K, the $|MR|_{6T}$ magnitude increases by ~24%, in the temperature range of $T^* < T < T_c$ with respect to its value at $T_c$. The data in this temperature regime is shown as blue circles. Whereas for $T < T^*$, the $|MR|_{6T}$ increases by ~96% on cooling from 160 to 10 K, which is four times of $\Delta MR$ in the FM regime. This reduced charge carrier scattering below $T^*$ could be ascribed to a new magnetic phase whose presence is also indicated by the magnetization and FMR measurements discussed earlier.

The changes in the electronic structure of the material as a result of this transition should lead to modifications in the carrier density, which can be probed through the measurements of the Hall effect. The Hall resistivity ($\rho_{xy}$) of the crystal has been measured over the same temperature and field ranges as the $M(H)$ and MR. The field scans of $\rho_{xy}$ shown in Fig. 2c reveal a clear non-linearity near zero field below $T_c$, a hallmark of the Anomalous Hall effect (AHE). The Hall resistivity of a ferromagnet is expressed as [38]

$$\rho_{xy} = \rho^H_{xy} + \rho^{AH}_{xy} = R_H B + \rho^{AH}_{xy}, \quad (2)$$

where $\rho^H_{xy}$ and $R_H$ are the normal Hall resistivity and Hall coefficient, respectively, whereas $\rho^{AH}_{xy}$ is the anomalous Hall resistivity. The Hall resistivity curves at $T = 10$ and 20 K appear to show an oscillatory character which are not intrinsic but arises due to a high noise floor at these temperatures. The details of data analysis for two representative temperatures, $T = 300$ and 40 K, are shown in Fig. 2d. Because of the AHE signal near zero field, the normal Hall coefficient ($R_H$) for each temperature is obtained from the linear slope of $\rho_{xy}$ vs $H$ at the high field region above ~2 T (denoted by red solid lines). Since the Hall resistivity in this regime is linear in field, we presume that it is devoid of any multiband effects, which generally lead to a non-linear field dependence of $\rho_{xy}$. Using the $R_H$ values at different temperatures and their polarity, we conclude a dominant hole carrier transport of carrier density in the range of $\approx(0.50\pm0.03)$ to $\approx(2.82\pm0.09)\times10^{21}$ cm$^{-3}$. These results are in agreement with an earlier report [24] on carrier density in F5GT. The variation of carrier density with temperature is shown in Fig. 2e (lower panel), which is unremarkable within the measurement uncertainty.

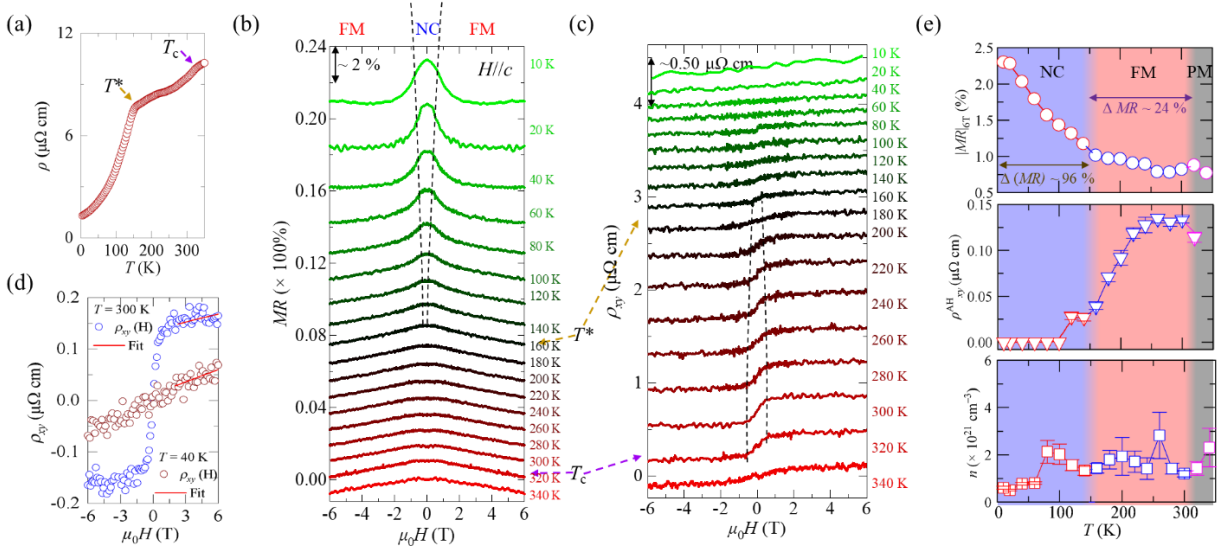

*Fig. 2. Magneto-transport properties of Fe₅GeTe₂.* (a) Temperature dependence of longitudinal resistivity (ρ) for the F5GT sample at zero magnetic field. (b) Field dependence of magnetoresistance MR = [ρ(H)- ρ(0)]/ρ(0) at different temperatures from T = 340 K to 10 K under H//c configuration. The black dashed lines are guides to the eyes, denoting critical fields for magnetization saturation seen in M(H) data of Fig. 2(b). The purple and gold dashed arrows represent $T_c$ and $T^*$, respectively. (c) Field dependence of Hall resistivity from T = 340 K to 10 K. The data in Figs. 2a and 2b are shifted for clarity. Therefore, the vertical axis labels are not the absolute values but provide the relative magnitudes. The dashed lines in Fig. 2c correspond to the saturation fields for out-of-plane magnetization in Figs. 1c and 1d. (d) The method to analyze Hall data is shown for two representative temperatures, T = 300 and 40 K. The Hall data for both temperatures in the high field region above 2 tesla are linear in field. (e) Temperature dependence of the magnitude of MR at 6 tesla ($|MR|_{6T}$, upper panel), anomalous Hall resistivity ($\rho^{AH}_{xy}$, middle panel) extracted by subtracting the normal Hall component from the total Hall effect data (Fig. 2c), and carrier density (n, lower panel). Note: Despite negative MR behavior as shown in Fig. 2b, only the absolute value of MR is used in this plot. The three different colored regions in Fig. 2e represent different magnetic phases.

Since the $\rho^{AH}_{xy}$ is directly proportional to magnetization [38], it is an effective probe of the magnetic state of the sample. The $\rho_{xy}$ data plotted in Fig. 2c show that while the $\rho^{AH}_{xy}$ is large close to the $T_c$, it starts to decrease upon cooling and eventually disappears at low temperatures. The temperature dependence of $\rho^{AH}_{xy}$ is presented in Fig. 2e (middle panel). The $\rho^{AH}_{xy}$ is determined by subtracting the normal Hall component ($\rho^H_{xy} = R_H H$) from the total Hall effect ($\rho_{xy}$) data. Interestingly, the AHE signal persists slightly above $T_c$, being observable at T = 320 K before vanishing at 340 K (Fig. 2c). The AHE above $T_c$ has been consistently reported in this family of compounds [10,24], presumably originating from the short-range magnetic correlations due to the quasi-2D layered structure that survives above the bulk $T_c$. Such behavior is commonly observed in quasi-two-dimensional magnets, where reduced dimensionality and magnetic fluctuations can stabilize short-range ordering well above the long-range transition temperature [39,40]. It is noted that the $\rho^{AH}_{xy}$ first increases as the temperature is reduced from 320 to 300 K and then remains nearly the same down to 260 K. In the temperature range of 240 to 160 K, a sharp drop in $\rho^{AH}_{xy}$ is seen. Finally, the AHE disappears below the FM–NC phase boundary temperature. The evolution of $\rho^{AH}_{xy}$ with temperature in F5GT is puzzling. Earlier studies have reported different temperature dependences of $\rho^{AH}_{xy}$ in bulk F5GT compound [10,41]. Both the suppression [10] and enhancement [41] of $\rho^{AH}_{xy}$ have been observed at temperatures $T \leq 120\text{-}160$ K, which are close to the $T^*$ in our sample. The different AHE behavior may be linked to variations in Fe vacancy concentration. As discussed above, our F5GT sample contains minimal vacancies. In contrast, earlier studies [10,41] have reported bulk $Fe_{5-x}GeTe_2$ samples with Fe vacancy $x \approx 0.3$ and 0.4. Magnetic properties of the FGT material system are highly sensitive to Fe content; in particular, the $T_c$ is strongly correlated with Fe vacancies and is markedly suppressed as vacancy concentration increases [36]. Such variations in Fe content modify the magnetic exchange interactions, which can alter magnetic properties and naturally lead to distinct AHE behavior. Therefore, sample-to-sample variations in AHE are expected in F5GT systems, explaining the discrepancy with previous reports. While systematic examination of the role of Fe-vacancy in setting the magnitude and sign of AHE would clarify this behavior, such a study lies beyond the scope of the present work but constitutes a promising direction for future research. The vanishing AHE below $T^*$ in the present case suggests the attenuation of FM interactions, which is consistent with the rise of in-plane and out-of-plane saturation fields below $T^*$ seen in the field-dependent isothermal magnetization measurements (Figs. 1c and d). In fact, the suppression of AHE at low temperatures is a common phenomenon in the FGT family. A similar effect has been reported in the sister compound $Fe_4GeTe_2$, where the AHE weakens at low temperature due to spin reorientation between in-plane and out-of-plane directions, while the magnitude of magnetization remains unchanged [8]. Moreover, similar attenuation of the AHE has been observed in other non-collinear spin systems, which likewise exhibit a pronounced deviation between the anomalous Hall response and the macroscopic magnetization. For example, $Mn_3Sn$ [42], $Co_3Sn_2S_2$ [43], $Mn_5Si_3$ [44], $Fe_3Sn_2$ [45], etc. The AHE in these materials is not solely dictated by net magnetization but is also governed by the underlying non-collinear spin structure [42–45]. This underscores that the microscopic spin texture also plays a decisive role in the anomalous Hall response beyond the contribution from the total magnetization. Therefore, our observation of the vanishing AHE below $T^*$ is consistent with the emergence of a canted, non-collinear magnetic state in F5GT. Hence, the unconventional evolution of MR and AHE with temperature in our F5GT sample manifests a possible magnetic phase transition below $T^*$.

To understand the nature of magnetic order below $T^*$, we measured the angular MR (AMR) at different temperatures under a fixed field of 6 tesla, as shown in Fig. 3. First, the $\theta$ dependence of MR is measured, as the magnetic field is rotated from out-of-plane ($\theta = 0°$) to in-plane ($\theta = 90°$) direction. The measurement setup is presented in Fig. 3a along with the behavior of $\theta_{AMR}$ defined as $\theta_{AMR} = (\rho(\theta)-\rho(0°))/\rho(0°)$ at several temperatures. For $T^* < T < T_c$ (i.e., for $T$ = 180, 220, and 260 K), the $\theta_{AMR}$ displays a weak two-fold anisotropy with the maximum and minimum at $\theta = 0°$ ($H//c$) and $\theta = 90°$ ($H//ab$), respectively, but with a negligible magnitude of ≈0.14% for $T$ = 260 K. Although weak, such $\cos\theta$-dependent $\theta_{AMR}$ can be attributed to the strong suppression of spin scattering when the field is aligned towards the easy axis, i.e., $H//ab$ ($\theta = 90°$). The $\theta_{AMR}$ magnitude is reduced to ≈0.10% on cooling from 260 to 220 K, and becomes nearly-isotropic at 180 K. On further cooling below $T^* \approx 160$ K, a dramatic reversal of the periodicity of $\theta_{AMR}$ is seen. Now, the maxima and minima of $\theta_{AMR}$ appear at $\theta = 90°$ and $\theta = 0°$, respectively. At first glance, such crossover in AMR anisotropy indicates a reorientation of the magnetic easy axis. However, our magnetization results demonstrate that the easy axis remains along or near the $ab$-plane for the entire temperature range (as discussed earlier). Thus, instead of a change in anisotropy axis, the scenario of magnetic phase transition near $T^*$ seems to be the likely reason for the change in the behavior of $\theta_{AMR}$.

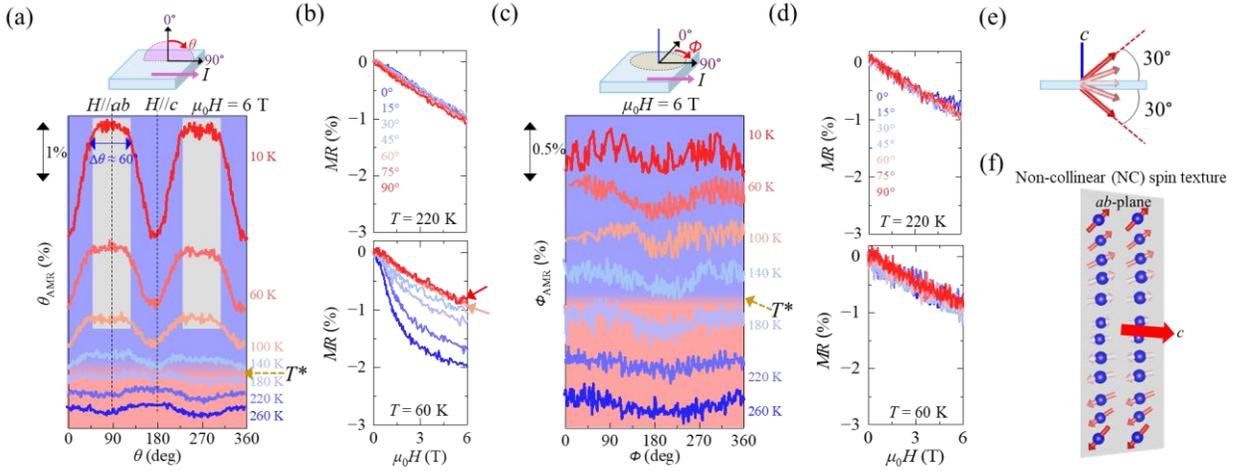

**Fig. 3. Angular dependence of resistivity of Fe$_5$GeTe$_2$.** (a) $\theta$-angle dependence of resistivity [$\rho(\theta)$]. The resistivity is normalized to zero-degree resistivity and termed as $\theta_{AMR} = [\rho(\theta)-\rho(0°)]/\rho(0°)$. Here, $\theta = 0°$ and $90°$ represent the magnetic field applied along the out-of-plane and in-plane directions, respectively. The shaded regime in Fig. 3(a) highlights the flat region in $\rho(\theta)$ measurements at low temperatures below $T^*$. (b) Field dependence of Normalized MR at $T$ = 220

*K (upper panel) and 60 K (lower panel) under different magnetic field orientations θ. The arrows in Fig. 3b (lower panel) highlight the near-isotropic MR within the angular range between θ = 60° and 90°. The measurement setup is presented in Fig. 3a. (c) ϕ-angle dependence of resistivity [ρ(ϕ)]. The resistivity is normalized to zero-degree resistivity and termed as $\phi_{AMR}$ = [ρ(ϕ)-ρ(0°)]/ρ(0°). Here, the magnetic field is applied along the in-plane directions for the entire angular sweep. Both θ and ϕ angular sweeps are measured above and below low-temperature magnetic anomaly ($T^*$ ≈ 160 K) from T = 260 K to 10 K at $\mu_0 H$ = 6 T magnetic field. (d) Field dependence of Normalized MR at T = 220 K (upper panel) and 60 K (lower panel) under different magnetic field orientations ϕ. The measurement setup is presented in Fig. 3c. Here, the MR in Figs. 3b and 3d are only shown for positive magnetic fields ranging from 0 T to 6 T. The pink arrows, denoting I in Figs. 3a and 3c, indicate the current direction. The different colored regions in Figs. 3a and 3c represent different magnetic phases. (e) Schematic showing moments canting towards the c-axis up to a maximum angle of θ ≈ ± 30° from the ab-plane. (f) Schematic illustration of a non-collinear spiral spin texture in the ab-plane in $Fe_5GeTe_2$. These schematics focus on the qualitative illustration of the overall non-collinear spin arrangement in $Fe_5GeTe_2$.*

Such a complex magnetism below $T^*$ is also highlighted by the plateau-like $\theta_{AMR}$ near θ = 90° (H//ab), as denoted by the grey shaded region in Fig. 3a. The AMR becomes isotropic within the angular range from θ ≈60° to θ ≈120°. The field dependence of MR for specific angles at 200 and 60 K is shown in Fig. 3b, upper and lower panels, respectively. At the higher temperature, the MR% is the same for all angles with a linear field dependence. However, at $T < T^*$, a non-linear field dependence of varying non-linearity emerges for angles θ ≤ 60 degrees. The plateau in AMR for θ between ≈ 60 to ≈ 120° signals an isotropic spin scattering. Notably, the plateau-like AMR has been seen in some AFM systems where it is attributed to spin-flop transition in a non-collinear spin configuration [46]. Although the system under consideration here is not an antiferromagnet, a scenario of canted moments on the *ab*-plane with a maximum canting angle of θ ≈ 30° above and below the *ab*-plane, as depicted in Fig. 3e, could be the reason behind the isotropic AMR within that angular range. Indeed, the moment-canting scenario at low temperature has been proposed for F5GT [30].

The non-collinear spin texture in F5GT is further manifested in the in-plane field rotation measurement of resistivity. Fig. 3c shows the angle-dependent MR measured by rotating the ab-plane of the sample in a 6 tesla in-plane field. For all the temperatures from $T$ = 260 to 10 K, the normalized $\phi$-dependent AMR [$\phi_{AMR} = (\rho(\phi)-\rho(0°))/\rho(0°)$] lacks field-orientation dependence. Compared to $\theta_{AMR}$, the magnetoresistance is nearly isotropic for in-plane field rotation, as shown in Fig. 3d. Such in-plane isotropic AMR is unexpected for a hexagonal crystal, where one would expect a six-fold symmetry in the $\phi_{AMR}$ measurement [46]. However, in systems like F5GT, where magnetic moments align in-plane, rotational symmetry is intrinsically broken, making crystal symmetry signatures harder to discern [47]. The absence of six-fold symmetry, even above $T^*$, also points to possible domain effects or twinning, which can disrupt the global symmetry expected from the lattice. A recent μ-ARPES study on the itinerant ferromagnet $Fe_3Sn_2$ demonstrated that perfect six-fold symmetry emerges only in a single crystallographic domain without in-plane ferromagnetism [47]. Similar high-resolution spectroscopic techniques with sufficient spatial resolution to resolve both magnetic and structural domains would be required to unambiguously separate crystal symmetry effects from domain/twinning influences in F5GT. Given these considerations, the unusual isotropic $\phi_{AMR}$ in our measurements may be linked to a non-collinear magnetic order within the *ab*-plane. In this scenario, where magnetic moments lack a preferred direction on the *ab*-plane, as sketched in Fig. 3f, the spin scattering may be insensitive to in-plane magnetic field orientation, causing an isotropic magnetotransport within the *ab*-plane. Thus, the AMR behavior for out-of-plane ($\theta$-) and in-plane ($\phi$-) field rotations in F5GT suggests a non-collinear spin texture in the *ab*-plane with possible moment canting towards the *c*-axis up to a maximum angle of $\theta \approx 30°$ from the *ab*-plane, as depicted in Figs. 3e and 3f. Such a non-collinear magnetic order predominantly confined within the *ab*-plane, with only a slight canting toward the out-of-plane direction, accounts for the absence of a low-temperature anomaly in $\chi_c$ under a 0.1 tesla field (Fig. 1a). A weak out-of-plane field of this magnitude is insufficient to perturb the canting and thus does not affect susceptibility. In contrast, a much stronger out-of-plane field ($\approx$ 1 tesla) enforces spin alignment along the *c*-axis (Fig. 1a). The observed non-collinear magnetic texture is also consistent with the suppression of the AHE below $T^*$ (Fig. 2c). The AHE is known to be strongly correlated with spin canting [48], and its attenuation can also occur within a non-collinear magnetic phase [49]. Future theoretical calculations are essential to clarify the temperature-driven evolution of the magnetic states and their interplay with the electronic transport in this system, which can be directly validated by experimental probes such as neutron diffraction and angle-resolved photoemission spectroscopy (ARPES).

**Physical origin and strategy to quench non-collinear spin order in Fe$_5$GeTe$_2$.**

The unusual magnetic properties of F5GT at low temperatures have been linked to changes in crystal symmetry driven by the ordering of Fe1 atoms between two split-sites located above or below the Ge atoms [23,30], as sketched in the inset of Fig. S1c [26]. At low temperatures, the formation of a Fe vacancy facilitates the swapping of the Fe1 positions between these two split-sites. An earlier STM study has shown that such vacancy-mediated ordering of Fe1 atoms stabilizes the ($\sqrt{3} \times \sqrt{3}$)$R30°$ supercell structure and consequently breaks the inversion symmetry of the crystal, giving rise to substantial antisymmetric DMI [23]. Here, we propose a competition between magnetic anisotropy and DMI as the origin of the low-temperature non-collinear spin texture in F5GT. Among the three FGT compounds, F5GT exhibits a weaker magnetocrystalline anisotropy ($K_m$). We calculated the $K_m$ using the equation $K_{eff} = K_m + K_{sh}$, where $K_{eff}$ and $K_{sh}$ are the effective uniaxial magnetic and shape anisotropies, respectively. The $K_{eff}$ is directly proportional to the saturation field $H_s$ along the hard axis (i.e., $c$-axis) and follows a relation $\mu_0 H_s = 2K_{eff}/M_{sat}$ [8], where $M_{sat}$ is the saturated magnetization. Using $\mu_0 H_s \approx 0.34$ T (at $T = 10$ K) and $M_{sat} \approx 2.1\mu_B$/Fe from our magnetization measurement, the $K_{eff}$ is estimated to be $\approx 0.14$ J/cm$^3$, which is close to the reported value of ~0.16 J/cm$^3$ for F5GT [24]. Furthermore, the $K_{sh}$ turns out to be $K_{sh} = -(1/2)\mu_0(M_{sat})^2 \approx -0.09$ J/cm$^3$. Using these values of $K_{eff}$ and $K_{sh}$, we estimate that the $K_m$ of F5GT is $\approx 0.23$ J/cm$^3$. The values of these parameters for our F5GT sample and the reported parameters from earlier works for Fe$_3$GeTe$_2$ [18,50,51] and Fe$_4$GeTe$_2$ [8] are summarized in Fig. 4a. The $K_{eff}$ and $K_m$ are reduced by increasing $N$ in Fe$_N$GeTe2 and attain the lowest value for F5GT.

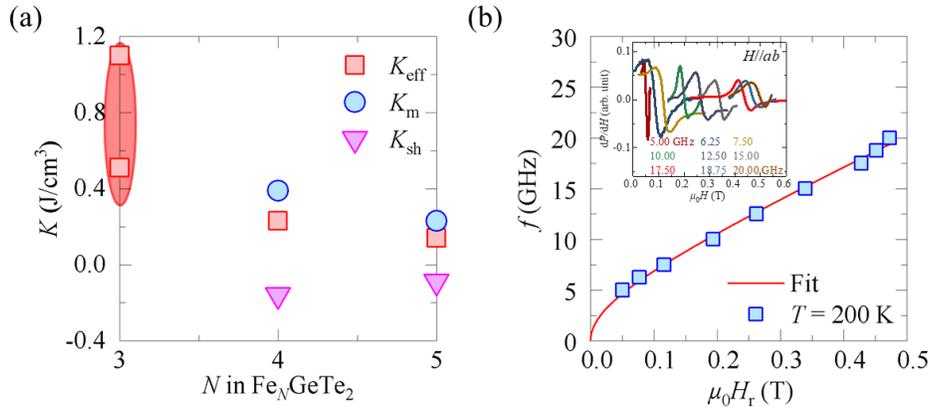

*Fig. 4. (a) Magnetic anisotropy parameters of Fe$_5$GeTe$_2$. Here, the uniaxial magnetic anisotropy ($K_{eff}$) is expressed as the sum of magnetocrystalline anisotropy ($K_m$) and shape anisotropy ($K_{sh}$) i.e., $K_{eff} = K_m + K_{sh}$. All the parameters for Fe$_3$GeTe$_2$ (N = 3) and Fe$_4$GeTe$_2$ (N = 4) are taken from Refs [18,50,51] and [8], respectively. (b) Input frequency (f) as a function of FMR field ($H_r$) at T = 200 K for Fe$_5$GeTe$_2$. The red plot indicates fitting of f vs $H_r$ to a fitting function $f = \gamma'\sqrt{[(H_r + 4\pi M_{eff})H_r]}$, where $\gamma' = |\gamma|/2\pi$, to determine the Lande g-factor using $g = |\gamma|\hbar/\mu_B$. The $H_r$ at different frequencies is obtained by fitting the experimental FMR spectra presented in the inset of Fig. 4(b) with the theoretical model given in equation 1. Inset: FMR spectra of a Fe$_5$GeTe$_2$ single crystal under an external DC magnetic field applied along the ab-plane at different GHz frequencies from 5 to 20 GHz at T = 200 K.*

The weak magnetocrystalline anisotropy of the F5GT sample can be understood based on the valence state of Fe ions. Both $Fe^{2+}$ and $Fe^{3+}$ ions can coexist in FGT compounds [52]. However, a recent theoretical work has demonstrated only a +3 valence state for Fe atoms in F5GT [52]. The $Fe^{3+}$ ion has five d-electrons ($d^5$), which means the Fe d-orbitals are half-filled, leading to the quenching of the orbital magnetic moment in F5GT. To confirm the orbital moment quenching, we determined the Landé g-factor (g) by fitting the frequency (f) dependence of the FMR resonance field ($H_r$) with the following equation (Fig. 4b)

$$f = \gamma'\sqrt{[(H_r + 4\pi M_{eff})H_r]}, \quad (3)$$

where $\gamma'$ is the reduced gyromagnetic ratio ($\gamma' = |\gamma|/2\pi$) and $M_{eff}$ is the effective magnetization. Here, we have performed broadband FMR measurement of F5GT single crystal at different input frequencies from f = 5 to 20 GHz at T = 200 K under an external DC magnetic field directed along the ab-plane of the sample (Inset: Fig. 4b). These FMR spectra are used to extract the $H_r$ of the resonance by fitting them with a two-component Lorentzian function given by the equation 1. Using a relation $g = |\gamma|\hbar/\mu_B$, we estimated $g = 2.1\pm0.1$, which is close to the spin-only Landé g-factor $g_s \approx 2$. This indicates a nearly zero orbital contribution to the magnetic moment, consistent with the negligible orbital moment compared to the spin moment calculated for F5GT [30]. Indeed, a recent FMR study has also confirmed the negligible orbital magnetic moment in F5GT [32]. Given a negligible orbital moment, the magnetocrystalline anisotropy in F5GT is expected to be very weak. On the other hand, a substantial DMI energy ranging from $(0.4–12)\times10^{-4}$ J/m$^2$ has been estimated in F5GT [23,41,53]. This is comparable to the known DMI values [$(1–30)\times10^{-4}$ J/m$^2$] for other magnetic materials such as $Co_2FeAl$ [54] and CoFeB [55], and found to be sufficient to induce the non-collinear magnetic texture in F5GT [53]. To clarify this, additional experimental and computational efforts are needed to directly quantify the DMI/$K_m$ ratio, providing a more rigorous basis for assessing the proposed competition between $K_m$ and DMI.

This mechanism is further elucidated by adopting an approach of quenching the low-temperature non-collinear spin texture in F5GT. The quenching task could be challenging given the complicated magnetic order that depends on the crystal symmetry, atomic ordering, magnetic anisotropy, and exchange interaction. Our work aims to shed light on possible strategies to suppress such non-collinearity and maintain the collinear magnetic order in vdW magnets. As discussed above, the key factor for weak $K_m$ in F5GT is the negligible orbital moment contribution due to the half-filled $d$-orbital of the $Fe^{3+}$ ion. Modifying the $d$-orbital occupancy of metal ions from the half-filled state may enhance the magnetic anisotropy and favor quenching of non-collinear magnetic order. However, enhancing $K_m$ alone may not be sufficient given the substantial DMI energy. For this, the DMI may also need to be attenuated. The inversion symmetry breaking caused by the complex Fe1 ordering is the reason behind DMI in F5GT. Therefore, maintaining the inversion symmetry may be necessary to suppress the non-collinear magnetic order. In addition, the weak magnetic exchange interactions are also responsible for the disruption of collinear FM correlations in F5GT [23]. The DMI in F5GT is almost ~10% of exchange interactions, and plays a primary role in forming a helical magnetic phase at low temperatures [23]. Therefore, strengthening magnetic exchange interactions could also facilitate the quenching of the non-collinear spin texture.

**Ni substitution in $Fe_5GeTe_2$.**

To realize the above scenario, we performed Ni-substitution at Fe sites to quench the non-collinear magnetic order in F5GT. An earlier work has also focused on Co substitution in F5GT, demonstrating its ability to tune the magnetic order effectively [57]. However, substituting Co for Fe changes the crystal structure and induces AFM order [57]. Substitution of Ni, on the other hand, preserves the FM ground state [27]. Further, Ni substitution in F5GT has been theoretically predicted [56] and experimentally confirmed [27] to display a site-selective doping behavior with preference for the Fe1 sites. To investigate the effect of Ni on the low-temperature magnetic order of F5GT, we synthesized $(Fe_{1-x}Ni_x)_5GeTe_2$, single crystals with $x$ = 0.15, 0.25, and 0.40, whose optical micrographs are presented in Fig. S1a [26]. The $(Fe_{1-x}Ni_x)_5GeTe_2$ crystals retain a crystal structure similar to that of F5GT as revealed by the high-intensity (00$L$) XRD peaks seen in Fig. S1b [26]. Further, the Bragg peaks systematically shift toward lower angles with the increasing Ni content, indicating an expansion of the interplanar spacing. This lattice expansion is quantified in Fig. S1c [26]. A detailed structure characterization in an earlier work has demonstrated the layer stacking change from ABC (space group $R\bar{3}m$) to AA stacking (space group $P\bar{3}m1$) above ~19% Ni substitution into F5GT [27]. These two types of stacking orders in F5GT share an identical intralayer structural motif but differ in their out-of-plane stacking sequence. As a result, the $c$-axis lattice parameter in the AA-stacking order is reduced to nearly one-third of that in the ABC-stacking configuration [19,27]. Indeed, the $c$-axis lattice parameter of $c \approx 9.72$ Å is obtained from the (00$L$) XRD spectra for the Ni-F5GT sample, which is close to the reported value [27] and nearly 1/3 of $c \approx 28.84$ Å for pristine F5GT. This modification of the crystal lattice indicates a transition from ABC- to AA-stacking upon substituting 40% Ni for Fe in F5GT.

Furthermore, the Ni substitution has substantial effects on the crystal lattice. Since the isotropic symmetric Fe-Fe exchange interactions primarily govern the $T_c$ in F5GT [56], Ni substitution is likely to enhance the $T_c$ because it optimizes the nearest-neighbor Fe–Fe distances and their effective coordination numbers, leading to a rearrangement of local Fe ordering that strengthens ferromagnetic interactions in Ni-F5GT. To clarify this, we have measured the temperature-dependent susceptibility ($\chi$) of the Ni- F5GT samples under $H//ab$ and $H//c$ fields of $\mu_0H = 0.1$ tesla (purple color) and 1 tesla (blue color), as shown in Fig. S3 [26] (for $x = 0.15$ and 0.25) and Fig. 5a (for $x = 0.40$). These measurements show a systematic rise in $T_c$ with Ni substitution. The composition dependence of $T_c$ is summarized in the supplementary material (Fig. S4) [26], together with the reported values for related compositions [27]. Our result agrees well with the reported results , confirming the successful Ni substitution in F5GT. It has been shown earlier that increasing the Ni concentration beyond ~ 36% leads to a drop in $T_c$ such that for ~ 86% Ni the crystal stays paramagnetic down to helium temperatures [27]. These findings indicate that while at lower concentrations Ni strengthens FM interactions, at the higher level of substitution, when Ni-Ni interactions become important, it promotes a paramagnetic ground state. Consequently, it is expected that the Ni substitution will affect the non-collinear state of F5GT.

Importantly, in contrast to pristine F5GT, no anomaly is seen in the low-temperature susceptibility of any Ni-substituted samples in either $\chi_{ab}$ or $\chi_c$ measurements performed at 0.1 and 1.0 tesla and shown in Fig. 5a and Fig. S3 [26]. In light of this result, the characteristic temperature $T^*$ is assigned only for the pristine F5GT sample in Fig. S4. The absence of $T^*$ in Ni-F5GT is also seen in the $\rho_{xx}(T)$ of this sample presented in Fig. 5b. Further, even in the sample with the lowest Ni concentration, there are no signatures of a $T^*$ in the $M(T)$ data down to 2.0 K. This highlights the sensitivity of non-collinear spin textures to Ni substitution in F5GT.

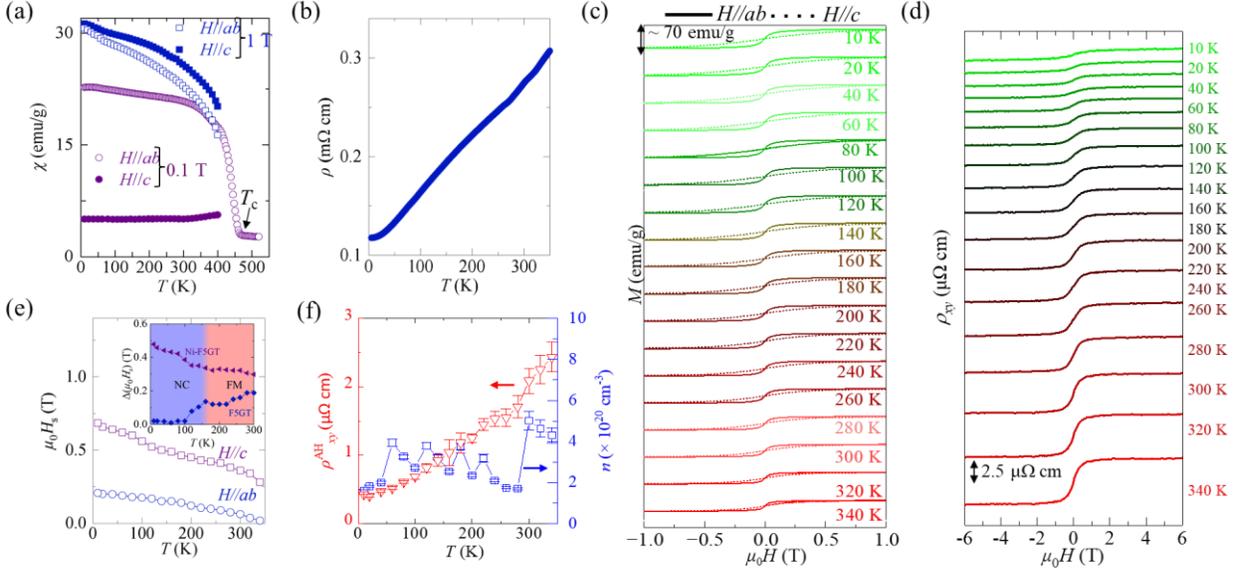

**Fig. 5. Magnetic properties of 40% Ni-substituted Fe$_5$GeTe$_2$ (Ni-F5GT).** *(a) Temperature dependence of susceptibility ($\chi$) under H//ab and H//c fields of $\mu_0H$ = 0.1 tesla (Purple color) and 1 tesla (Blue color) under a ZFC protocol. The black arrow denotes the PM to FM transition temperature $T_c \approx 472$ K. (b) Temperature dependence of longitudinal resistivity ($\rho$) of the same sample in zero magnetic field. (c) M(H) of the sample under in-plane (solid line) and out-of-plane (dashed line) fields from T = 340 to 10 K. (d) $\rho_{xy}(H)$ data of Ni-F5GT at several temperatures from 340 to 10 K. (e) Variation of $(H_s)_{ab}$ and $(H_s)_c$ critical fields as a function of temperature. Inset: Temperature dependence of the anisotropy between $(H_s)_{ab}$ and $(H_s)_c$ for F5GT and Ni-F5GT, which is given by $\Delta|H_s| = (H_s)_c - (H_s)_{ab}$. Magnetic phases of the sample have been identified with different colors. (f) Temperature dependence of anomalous Hall resistivity ($\rho^{AH}_{xy}$; left vertical axis) extracted by subtracting the normal Hall component from the total Hall effect data [Fig. 5(d)] and the carrier concentration (n; right vertical axis).*

The lack of low-temperature anomaly in the Ni-F5GT crystal is also supported by the field dependence of magnetization $M$(H) measured from 340 to 10 K under $H//ab$ (solid line) and $H//c$ (dashed line), as seen in Fig. 5c. The $M$(H) below $T_c$ displays field-driven saturation, which is expected in ferromagnets. Alongside the tuning of $T_c$, Ni substitution also reduces the $M_{sat}$ by a factor of ~1.6 (Figs. 1c and 5c). As discussed in the previous section, Fe atoms in F5GT may adopt the +3 valence state, resulting in half-filled d-orbitals with $3d^5$ ($S^{5/2}$) electronic configuration, which maximizes the net magnetic moment. Replacing $Fe^{3+}$ with $Ni^{3+}$ ($3d^7$) or $Ni^{2+}$ ($3d^8$) ions with the spin–only moment of 3/2 and 1, respectively, will decrease the net moment. Furthermore, the modulation of $M_{sat}$ with Ni substitution is accompanied by the tuning of $H_s$. Contrary to the non-monotonic temperature dependence of $H_s$ for pristine F5GT (Fig. 1d), the Ni-F5GT sample exhibits a systematic increase in $H_s$ upon cooling (Fig. 5e), which is typical for collinear ferromagnets. The anisotropy between $(H_s)_{ab}$ and $(H_s)_c$, given by $\Delta|H_s| = (H_s)_c - (H_s)_{ab}$, does not show any dramatic transition with varying temperature for Ni-F5GT. On the other side, as discussed earlier, the F5GT sample exhibits nearly isotropic magnetic order, with a significant suppression of $\Delta|H_s|$ below $T^*$ (Inset: Fig. 5e). This behavior can be attributed to the onset of non-collinear magnetic order, where moments start to cant relative to each other within the ab-plane and slightly along the c-axis, as illustrated in Figs. 3e and 3f, which smears out the directional preference of the spins, effectively lowering the magnetic anisotropy below $T^*$. Thus, the low-temperature magnetic phase transition seen in pristine F5GT is not apparent in the magnetization measurements of the Ni-F5GT sample. This is also illustrated by the evolution of AHE with temperature. Fig. 5d presents the Hall data of the Ni-F5GT crystal from 10 to 340 K. The $R_H$ values at different temperatures and their polarity indicate that the carrier transport is dominated by hole carriers, consistent with the pure F5GT crystal (Fig. 2e). However, the carrier density calculated from these data ranges between $\approx(5.01\pm0.44)\times10^{20}$ and $\approx(1.58\pm0.05)\times10^{20}$ cm$^{-3}$ across all measured temperatures (Fig. 5f; right vertical axis), which is about an order of magnitude smaller than in pristine F5GT. This reduction indicates electron doping induced by Ni substitution, as reported previously [27]. While the carrier density does not vary significantly with temperature, the $\rho^{AH}_{xy}$ exhibits a monotonic reduction upon cooling (Fig. 5f; left vertical axis). The signature of AHE near the zero-field regime and the corresponding $\rho^{AH}_{xy}$ is apparent down to $T = 10$ K, as seen in Figs. 5d and 5e (right panel), which sharply contrasts with the disappearance of AHE below $T^*$ in F5GT. This clearly shows a robust FM state down to 10 K upon Ni substitution. In addition, the absence of a magnetic phase transition is also revealed by the monotonic $\theta$ dependence of MR [$\theta_{AMR} = (\rho(\theta)-\rho(0°))/\rho(0°)$], which is shown in Fig. 6a for the measurements performed at 6 tesla in the temperature range of $T = 10$ to 260 K. Here the $\theta_{AMR}$ displays a normal two-fold anisotropy with the maxima and minima at $\theta = 90°$ ($H//ab$) and $\theta = 0°$ ($H//c$), respectively over the entire temperature range. Furthermore, the field dependence of MR for both $T = 220$ and 60 K is negative and nearly the same for all $\theta$-angles (Fig. 6b).

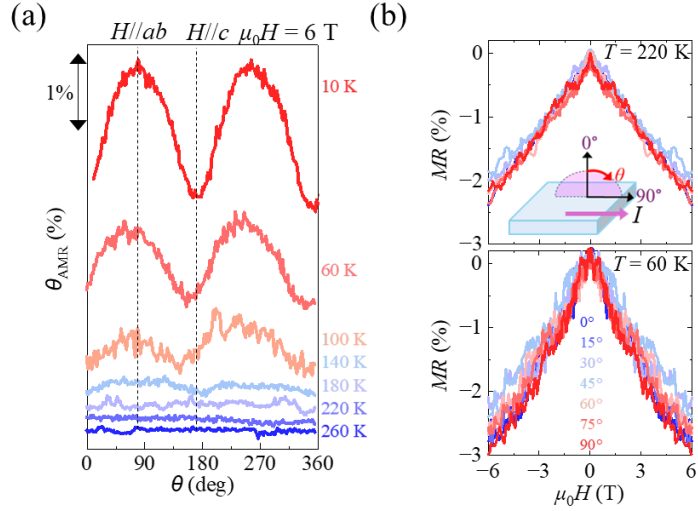

*Fig. 6. Angular dependence of resistivity of Ni-F5GT. (a) $\theta$-angle dependence of resistivity [$\rho(\theta)$]. The resistivity is normalized to zero-degree resistivity and termed as $\theta_{AMR} = [\rho(\theta)-\rho(0°)]/\rho(0°)$. (b) Field dependence of Normalized MR at $T = 220$ K (upper panels) and 60 K (lower panels) under different magnetic field orientations. The measurement setup is presented in Fig. 3b. Here, $\theta = 0°$ and $90°$ represent field directions along the c-axis and along the ab-plane, respectively.*

The results presented above demonstrate the quenching of the low-$T$ non-collinear spin texture phase in F5GT on the substitution of Ni at the Fe sites. Such suppression of non-collinearity implies reduced DMI compared to the Ni-free crystals of F5GT. As discussed earlier, Ni substitution has significant impacts on the crystal structure. However, a recent theoretical work predicts the existence of ($\sqrt{3} \times \sqrt{3}$) superstructures in Ni-F5GT in a manner similar to F5GT [58]. This may imply that the inversion symmetry is also broken in Ni-F5GT. If this is the case, a drastic suppression of DMI may not occur. Instead, the quenching of non-collinear magnetic order in Ni-F5GT may be attributed to the enhanced FM interactions and magnetic anisotropy. The substantial rise in $T_c$ with Ni substitution indicates stronger FM exchange interactions, which may overshadow the DMI, thus suppressing the low-$T$ non-collinear magnetic phase. Furthermore, the Ni substitution may also strengthen magnetocrystalline anisotropy because of the $d^7$ and $d^8$ configurations of the $Ni^{3+}$ and $Ni^{2+}$ ions, respectively, where orbital angular momentum may be partially quenched. In fact, a greater anisotropy between in-plane and out-of-plane saturation fields for Ni-F5GT compared to the pristine F5GT (Inset: Fig. 5e, left panel) is suggestive of enhanced magnetocrystalline anisotropy.

## IV. CONCLUSION

In conclusion, we performed a detailed study of the low-temperature magnetism of a high-$T_c$ vdW ferromagnet $Fe_5GeTe_2$. Our magnetization, FMR, and electrical transport measurements reveal a possible magnetic phase transition from FM to a complex non-collinear magnetic state below $T^* \approx 160$ K. Such a rich magnetic order at low temperatures might arise from the competition between magnetocrystalline anisotropy and DMI. It has been shown that this competition can be marginalized by 15% Ni substitution for Fe in Fe5GT, which presumably enhances the anisotropy and FM interactions. Furthermore, this work introduces a strategy to quench the non-collinear spin texture, which can be generalized to other related materials. Stabilizing a collinear magnetic order can offer significant technological benefits, including enhanced robustness of magnetic states, improved reproducibility in spintronic devices, and more reliable control of switching properties in vdW magnets.

**Acknowledgements**

This work has been funded by the United States Department of Defense (DoD) under grant No. W911NF2120213.